\documentclass{svjour2}                    
\smartqed  
\usepackage{graphicx}
\usepackage{mathptmx}      
\usepackage{amsmath,verbatim}
\usepackage[pdftex,usenames]{color}
\usepackage{hyperref}
\usepackage{latexsym}
\newcommand{\av}[1]{\langle {#1} \rangle}

\newcommand{\ep}{\varepsilon}
\newcommand{\eps}{\varepsilon}

\journalname{J Stat Phys}
\begin{document}

\title{Mean-field analysis of the $q$-voter model on networks
}


\author{Paolo Moretti     \and
        Suyu Liu         \and
        Claudio Castellano   \and
        Romualdo Pastor-Satorras
}

\authorrunning{S. Liu, P. Moretti, C. Castellano, and R. Pastor-Satorras} 

\institute{Paolo Moretti  \at
           Departamento de Electromagnetismo y F\'isica de la Materia and
           Instituto ``Carlos I'' de F\'isica Te\'orica y Computacional, 
           Universidad de Granada,
           Facultad de Ciencias,
           Fuentenueva s/n, E-18071 Granada,
           Spain
           \and
           Suyu Liu  \at
           State Key Laboratory of Industrial Control Technology
           Institute of Cyber-System and Control
           Zhejiang University
           Hangzhou 310027, P.R.China
           \and
           Claudio Castellano   \at
           Istituto dei Sistemi Complessi (ISC-CNR), via dei Taurini
           19, I-00185 Roma, Italy \at
           Dipartimento di Fisica, ``Sapienza'' Universit\`a di
           Roma, P.le A. Moro 2, I-00185 Roma, Italy
           \and
           Romualdo Pastor-Satorras \at 
           Departament de F\'\i sica i Enginyeria Nuclear,
           Universitat Polit\`ecnica
           de Catalunya, Campus Nord B4, 08034
           Barcelona, Spain
}

\date{Received: date / Accepted: date}

\maketitle

\begin{abstract}
We present a detailed investigation of the behavior of the nonlinear
$q$-voter model for opinion dynamics. At the mean-field level we
derive analytically, for any value of the number
$q$ of agents involved in the elementary update, the phase diagram,
the exit probability and the consensus time at the transition point.
The mean-field formalism is extended to the case that the interaction pattern
is given by generic heterogeneous networks. We finally discuss the case
of random regular networks and compare analytical results with 
simulations.

\end{abstract}

\section{Introduction}
\label{intro}

Ordering dynamics is a classical subject of nonequilibrium statistical
physics, regarding phenomena such as magnetic ordering or
phase-separation of mixtures~\cite{Bray94,KineticViewRedner}.  Also in
the biological and social domains the question of how order emerges
and evolves over time is of fundamental interest.  This has recently
led to the introduction of a wealth of models for ordering dynamics in
interdisciplinary settings, inspired by the physical paradigms but
exhibiting a diversity of specific distinctive
features~\cite{Castellano09}.  In this context the voter
model~\cite{Clifford73,Holley:1975fk} plays a prominent role for
several reasons: it is extremely simple; it has a natural
interpretation in several fields (ecology, genetics, opinion
dynamics); it can be exactly solved in regular lattices for any number
of spatial dimensions~\cite{liggett99:_stoch_inter,KineticViewRedner}.

The voter model is defined as follows: A population of agents is
placed on a regular lattice or a graph of size $N$. Each agent
is endowed with a binary spin variable, representing two opposed
opinions, and taking values $\sigma=\pm1$. At each time step $t$,
an agent $i$, together with one nearest neighbor $j$, are selected at
random, and the state of the system is updated as $\sigma_i :=
\sigma_j$, the first agent copying the opinion of its
neighbor. Time is updated $t \to t + 1/N$. In this way, the
flipping probability, i.e. the probability that a spin surrounded by a
fraction $x$ of spins in the opposite state flips, is $f(x)=x$.
Starting from a disordered initial state, this dynamics leads in
finite systems to a uniform state with all individuals sharing the
same opinion (the so-called consensus).  From the social dynamics
point of view, interest is mainly focused on the exit probability
$E(x)$ and the consensus time $T_N(x)$, defined as the probability
that the final state corresponds to all agents in the state $+1$ and
the average time needed to reach consensus in a system of size $N$,
respectively, when starting from a homogeneous initial condition with
a fraction $x$ of agents in state $+1$~\cite{Castellano09}.  On
regular lattices voter dynamics conserves on average the global
magnetization $m = \sum_{i=1}^{N} \sigma_i/N$, and this implies $E(x)
= x$.  On the other hand, for $d\geq2$~\cite{1751-8121-43-38-385003}
the consensus time takes the form
\begin{equation}
  \label{eq:1}
  T_N(x) = - N_\mathrm{eff} \left[ x \ln (x) + (1-x) \ln (1-x)
  \right], 
\end{equation}
with $N_\mathrm{eff} \sim N^2$ in $d=1$, $N_\mathrm{eff} \sim N \log N$ in
$d=2$, and $N_\mathrm{eff} \sim N$ in $d>2$
(the latter being also the mean-field result)~\cite{KineticViewRedner}. 

In the social context voter dynamics becomes all the more relevant
when it is considered on top of complex networks, which are more
realistic representations of social contact or relationship patterns
\cite{barabasi02,Newman2010}.  The analysis on these substrates
reveals nontrivial differences with respect to lattices.  For example,
now the order in which interacting individuals are selected
matters~\cite{Suchecki05,castellano05:_effec}.  Moreover, relevant
quantities such as the exit probability and the consensus time turn
out to depend on the heterogeneity of the contact pattern, as measured
by the degree distribution $P(k)$
\cite{PhysRevLett.94.178701,Sood08,Pugliese09}.  On complex networks
the main conserved quantity is not the total magnetization, but a
connectivity weighted magnetization, $\omega$.  Thus, it is possible
to show that the exit probability is equal to this conserved quantity,
while the consensus time takes the form \cite{PhysRevLett.94.178701}
\begin{equation}
 \label{eq:26}
 T_N(\omega) = - N_\mathrm{eff} \left[ \omega \ln (\omega) +
   (1-\omega) \ln (1-\omega) 
 \right], 
\end{equation}
where now $N_\mathrm{eff}= N \av{k}^2/\av{k^2}$, with $\av{k^n} =
\sum_k k^n P(k)$ the $n$-th moment of the degree distribution.

In the case of regular lattices, the simple behavior of the voter
model has been shown to be more generic than its original definition
apparently shows.  Early work showed that the voter
model lies at a transition point between a ferromagnetic ordered phase
and a paramagnetic disordered phase, such that any small perturbation
is able to drive it out of the transition and radically change its
behavior \cite{Deoliveira93,Drouffe99,Molofsky99}. Further work on
this issue, based on numerical simulations and phenomenological field
theories has finally led to the realization that there exists a whole
generalized voter (GV) universality class encompassing models at an
order-disorder transition driven by interfacial noise between
``dynamically equivalent'' absorbing states. The equivalence can be
enforced either by $Z_2$-symmetric local rules, or by global
conservation of the magnetization~\cite{Dornic01,AlHammal05}. The
newly observed GV universality class is defined 
in terms of the critical exponents characterizing the
transition. However, recent research has shown that other
characteristics, such as the shapes of the exit probability and the
consensus times, exhibit nonuniversal features, depending on the
microscopic details defining the model~\cite{Castellano12}.

The investigation of the existence and the properties of the GV class
is based on the consideration of \textit{nonlinear} voter models, in
which the flipping probability $f(x)$ is a function taking a nonlinear
form \cite{Dornic01,Vazquez08}. In the general case, the nonlinearity
of the flipping probability can be considered to be modulated by a
tunable parameter $\eps$ and the GV behavior occurs for a critical
value $\eps_c$, {that separates a ferromagnetic (ordered) phase,
  in which the systems orders with a time that scales logarithmically
  with the system size, from a paramagnetic (disordered) phase, in
  which the systems orders at exponentially large times.} Few exact
results are available about these models, since contrary to the
standard voter model, nonlinear voter models are not in general
analytically solvable. Therefore, our understanding of the GV class is
mainly based on numerical simulations and on the analysis of field
theories, which are studied also numerically or by means of
renormalization group arguments~\cite{Canet05}. In this sense,
analytical results for nonlinear voter models can in general be
produced only at the mean field level.

In this paper we focus in the analysis of the mean field behavior on
networks of a nonlinear voter model recently introduced in the context
of opinion dynamics the $q$-voter model
\cite{PhysRevE.80.041129}. This model is characterized by two
parameters, representing the number of interacting voters $q$, and a
source of noise $\eps$. A critical value of the parameter $\eps$ 
identifies the GV transition,
taking a functional form depending on the particular value of
$q$. Considering initially the simplest case of a fully connected
graph, we solve the Fokker-Planck equation that represents the
dynamics, and obtain analytical information for different values of
$q$. We show that, even in the simplest mean-field case, nonlinear
voter models can show a very rich behavior, characterized by a complex
phase diagram for sufficiently large values of $q$. We then introduce
a theoretical treatment, within the heterogeneous mean-field
framework, for the $q$-voter dynamics on generic uncorrelated
networks. We finally turn to the consideration of random regular
networks as substrate of the dynamics.

\section{The $q$-voter model}
\label{sec:q-voter-model}

On an arbitrary lattice or graph, the $q$-voter model is defined as
follows: Each node hosts a spin, with a binary value $\sigma_i = \pm
1$. The system evolves dynamically by selecting, at a given time $t$,
a randomly chosen spin located at, say, node $i$, with state
$\sigma_i$. Additionally, $q$ neighbors of node $i$ are also chosen at
random, allowing for repetition.  If all the $q$ neighbors are in the
same state, $\sigma_q$, the original spin takes the value of the $q$
neighbors, $\sigma_i := \sigma_q$; otherwise, and with an independent
probability $\ep$.the original spin flips its state, $\sigma_i :=
-\sigma_i$. In any case, time is updated $t \to t + 1/N$, where $N$ is
the total number of sites.

It is easy to see that this model is non-linear. Consider the
probability $f$ that a site flips as a function of the fraction $x$ of
neighbors in the opposite state. The flipping probability can be
easily seen to take the form
\begin{equation}
  f(x,\eps, q) = x^q + \ep \left[1-x^q - \left(1-x\right)^q\right].
  \label{f}
\end{equation}
For $q=1$, we recover, for any value of $\ep$, the standard voter
model, namely $f(x,\eps, 1)= x$.  For $q=2$, on the other hand, the flipping
probability takes the form
\begin{equation}
  \label{eq:17}
  f(x,\eps, 2) = 2 \eps x+(1-2 \eps) x^2.
\end{equation}
For $\eps=1/2$, $f(x, 1/2, 2) = x$, and therefore linear voter
behavior must be observed for this combination of parameters in any
kind of substrate.
In order to observe truly nonlinear behavior in the $q$-voter model
(in the sense that $f(x) \neq x$ at the transition),
we must thus consider values of $q$ larger than $2$. 

\section{The $q$-voter model on fully connected graphs}
\label{sec:mean-field-theory}

The case of a fully connected graph corresponds to the classic
mean-field behavior, in which every node is neighbor of all the other
nodes in the system. Therefore, choosing $q$ neighbors at random
corresponds to the random selection of any $q$ nodes in the system.
In this case, the state of the system is fully characterized by number
$n$ of sites in state $+1$. {The behavior of the $q$-voter model
  can be derived in this regime by means of the exact diffusion
  formalism developed in Ref.~\cite{1751-8121-43-38-385003}. Here we
  follow however a more intuitive approach based on the Fokker-Planck
  equation \cite{Gardinerbook,blythe07:_stoch_model}}. The probability
of finding the systems in state $n$ at time $t$ fulfills the master
equation
\begin{equation}
  \frac{\partial P(n, t)}{\partial t} = 
  \sum_{n' \neq n} T(n|n') P(n',t) - \sum_{n' \neq n} T(n'|n) P(n,t),
\label{eq:12}
\end{equation}
where $T(n|n')$ are the transition rates from a state with $n'$
to a state with $n$ positive spins. The only non-zero transition rates
can be seen to take the form, for a fully connected graph,
\begin{eqnarray}
  T(n+1|n) &=& \frac{1}{\Delta} (1-x)  f(x,\ep, q),\\
  T(n-1|n) &=& \frac{1}{\Delta} x  f(1-x,\ep ,q),\\
  T(n|n) &=& \frac{1}{\Delta} \left[1 - (1-x)  f(x,\ep, q) - x
    f(1-x,\ep, q)\right],
\end{eqnarray}
where $\Delta=1/N$ is the elementary time scale of the problem and we
have defined $x=n/N$. Performing an expansion up to second order in
$\Delta$, the previous master equation can be converted into a
simplified Fokker-Planck equation, valid in the limit of large system sizes,
taking the form
\begin{equation}
  \frac{\partial P(x, t)}{\partial t} =
  - \frac{\partial}{\partial x} \left[  v(x)  P(x, t)\right] + 
  \frac{1}{2} \frac{\partial^2}{\partial x^2}  \left[  D(x)  P(x,
    t)\right],  
  \label{eq:14}
\end{equation}
where we have defined the drift $v(x)$ and the diffusion
coefficients $D(x)$, taking the forms
\begin{eqnarray}
  v(x) &=& (1-x) f(x,\eps,q) - x f(1-x,\eps,
  q), \label{drift} \\ 
  D(x)  &=& \frac{1}{N} \left[(1-x) f(x,\eps,q) + x
    f(1-x,\eps,q)\right].\label{diffusion} 
\end{eqnarray}
From the Fokker-Planck equation, application of a standard formalism
\cite{Gardinerbook} leads to an exit probability $E(x)$ satisfying the
differential equation
\begin{equation}
  v(x) \partial_x E(x) +
  \frac{1}{2}D(x) \partial^2_x E(x) = 
  0, \label{eq:1xxx} 
\end{equation}
with boundary conditions $E(0)=0$ and $E(1)=1$, while the average
time until consensus, $T_N(x)$, is given by
\begin{equation}
   v(x) \partial_x T_N(x)  + \frac{1}{2}
   D(x)  \partial^2_x T_N(x) =  
   -1,\label{eq:6} 
\end{equation}
with boundary conditions $T_N(0)= T_N(1)=0$. 

Linear flipping probability $f(x)=x$ implies absence of drift,
$v(x)=0$ and then linear exit probability $E(x)=x$.
Moreover, it is easy to see from Eqs.~\eqref{eq:1}
and~\eqref{eq:6}, that a zero drift is necessary and sufficient
condition to yield, for any diffusion $D(x)$ scaling as
$1/N$, $E(x)=x$ and $T_N(1/2) \sim N$, which are the two main
signatures of voter behavior at the mean-field level. Let us consider
what is the behavior of the $q$-voter model for different values of
$q$.

\subsection{Case $q=2$}
For $q=2$, the drift and the diffusion coefficient take the form
\begin{eqnarray}
  \label{eq:5}
  v(x) &=& -(1 - 2 \ep) (1 - x) x (1 - 2 x),\\
  D(x) &=& (1 + 2 \ep) (1 - x) x/N.
\end{eqnarray}
The value $\eps=1/2$ leads to linear voter behavior, with
$v(x)=0$. On the other hand, the drift takes the form
$D(x)=2x(1-x)/N$, leading to a consensus time of the form
\begin{equation}
  \label{eq:16}
  T_N(x) = - N \left[ x \log x + (1-x) \log(1-x) \right],
\end{equation}
which coincides with the form observed in regular lattices for
dimension $d \geq 2$, Eq.~\eqref{eq:1}. From Eq.~\eqref{eq:5}, we can
easily see that 
$\eps<1/2$ implies $v(x)<0$ for $x<1/2$ and $v(x)>0$
for $x>1/2$, which is indicative of a ferromagnetic ordered state. The
opposite behavior is found for $\ep>1/2$, indicating a disordered
paramagnetic phase. Therefore, the point $\eps=1/2$, corresponding to
linear voter behavior, lies at a transition between a ferromagnetic
and a paramagnetic phase.

\subsection{Case $q=3$}

In the case $q=3$, we have
\begin{eqnarray}
  \label{eq:7}
  v(x) &=& (-1 + 3 \eps) (1 - x) x (1 - 2 x),\\
  D(x) &=& (1 - x) x [1 + 3 \eps - 2 x(1-x)]/N.
  \label{eq:19}
\end{eqnarray}
The value $\eps=1/3$ separates again a ferromagnetic phase at
$\ep<1/3$ from a paramagnetic phase for $\eps>1/3$. Exactly at
$\eps=1/3$, we recover $v(x)=0$, which leads to a linear exit
probability, but now $D(x)=2 (1 - x) x (1 - x + x^2)/N$.  The
diffusion term is now different from the case of the linear voter, and
as a consequence it will lead to a different form of the consensus
time as a function of $x$. The form can be recovered by integrating
Eq.~(\ref{eq:6}), upon substituting the diffusion in
Eq.~\eqref{eq:19}, namely
\begin{equation}
  \label{eq:8}
  (1 - x) x (1 - x + x^2) \partial_x^2 T_N(x) = -N.
\end{equation}
The integration of this differential equation takes the form 
\begin{eqnarray}
  T_N(x)=&-&N\left[ x \ln x + (1-x)\ln(1-x)
    -  \frac{1}{2}\ln(1-x+x^2) \right. \nonumber \\
  &+& \left.
    \frac{1-2x}{\sqrt{3}}
    \tan^{-1}\frac{1-2x}{\sqrt{3}} - \frac{\pi}{6 \sqrt{3}} \right] 
  \label{eq:9}
\end{eqnarray}

\begin{figure}[t]
  \begin{center}
    \includegraphics [width=0.7\textwidth]{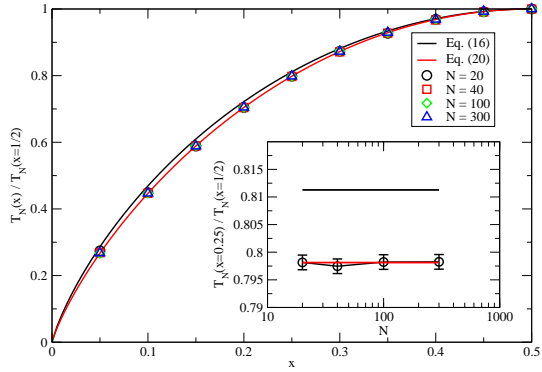}
    \caption{(color online) Main: Normalized consensus time for the
      $q$-voter model on the fully connected graph for $q=3$ at the
      critical point $\eps=1/3$.  For comparison, the entropic form,
      Eq.~\eqref{eq:16}, and the one given by Eq.~(\ref{eq:9}) are
      also plotted.  Inset: Normalized consensus time for $x=0.25$ as
      a function of $N$, showing the perfect agreement with the
      prediction of Eq.~(\ref{eq:9}). {Reported results from
        numerical simulations, averaged over $10^4$ independent
        realizations.}}
      \label{fig:Tvxq=3}
  \end{center}
\end{figure}

In Fig.~\ref{fig:Tvxq=3} we plot the normalized consensus time
$T_N(x)/T_N(1/2)$ for the $q=3$-voter model and $\ep=1/3$,
Eq.~(\ref{eq:9}), compared with the result from the linear voter
model, Eq.~\eqref{eq:16}. The difference induced by the nonlinear form
of $f(x,1/3,3)$ is small but noticeable, and indeed it is recovered by
means of numerical simulations on a fully connected graph (symbols in
Fig.~\ref{fig:Tvxq=3}).

\subsection{Case $q=4$}
\label{sec:case-q=4}

In the more interesting case of $q=4$, the drift and diffusion
coefficient take the forms
\begin{eqnarray}
  v(x) &=& (1 - x) x (1 - 2 x)  [-1 + 4 \ep +x(1-x)(1-2\ep)],\\
  D(x) &=&(1 - x) x [1 + 4 \eps - (3 + 2 \eps)x(1-x)]/N.
\end{eqnarray}
From this form of the drift, it is easy to check that there exists no
value of $\eps$ that allows it to vanish for every value of
$x$. Therefore, there is no region of linear voter behavior in the
phase diagram of this model. In order to visualize the form of this
phase diagram, it is convenient to change variables and represent the
drift as a function of the average magnetization $\phi=2x-1$. With
this transformation, the drift takes the form
\begin{equation}
  \label{eq:23}
   v(\phi) = \frac{1}{16} \left(1-\phi ^2\right) \phi
    \left((1-2 
   \varepsilon) \phi ^2+3-14 \varepsilon \right).
\end{equation}
A ferromagnetic phase will correspond to the region of values of 
$\eps$ for which  $\phi v(\phi) > 0$ for all $\phi$, while a
paramagnetic phase corresponds to  $\phi v(\phi) < 0$. From
Eq.~\eqref{eq:23}, we can see that the nature of the phase is given by
the sign of the function
\begin{equation}
  \label{eq:31}
  F_\eps(\phi)= 3 -14\eps +(1-2\eps)\phi^2.
\end{equation}
The ferromagnetic phase corresponds to $F_\eps(\phi)>0$, which takes
place in the region $\eps<3/14$; the region $\eps>1/4$ corresponds to
a paramagnetic phase with $F_\eps(\phi)<0$; on the other hand, the
region $3/14 < \eps < 1/4$ is a mixed phase, which is ferromagnetic or
paramagnetic depending on the initial value of $\phi$. In particular,
initial conditions with magnetization $\phi^2 <
\frac{3-14\ep}{2\ep-1}$ lead to a paramagnetic behavior, while $\phi^2
> \frac{3-14\ep}{2\ep-1}$ corresponds to a ferromagnetic phase.

\begin{figure}[t]
  \centering
  \includegraphics [width=0.7\textwidth]{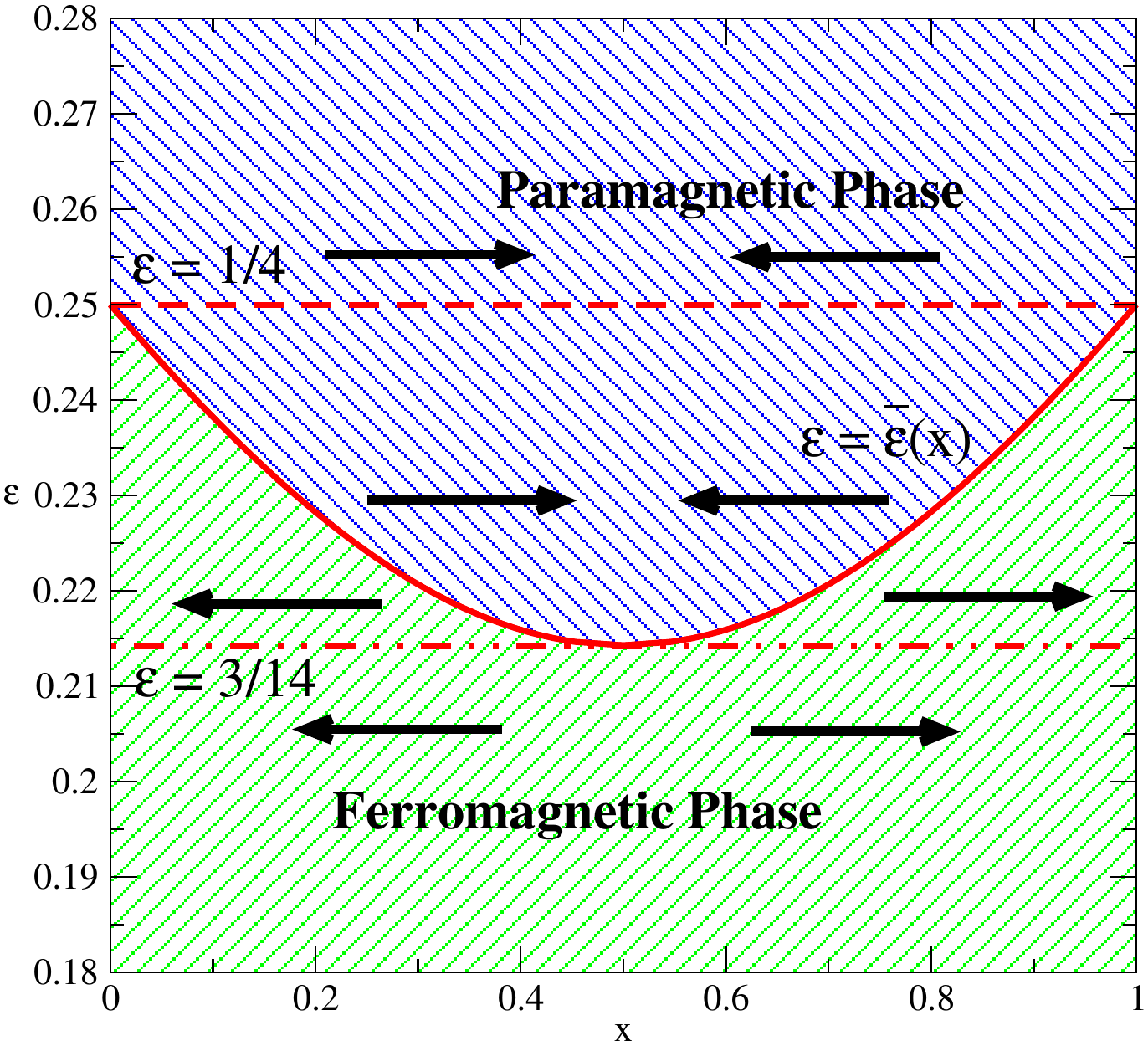}
  \caption{(color online) Phase diagram for the $q=4$-voter model on
    fully connected networks. {The arrows indicate the direction
      of the drift for different values of $x$}.}
  \label{fig:phasediag}
\end{figure}

The ensuing phase diagram for the $q=4$-voter model on fully connected
graphs is depicted, as a function of the density $x$ of $+1$ spins, in
Fig.~\ref{fig:phasediag}.
Summarizing its structure, for $\ep>1/4$, there is a paramagnetic
disordered phase, with $T_N(x)$ growing exponentially with $N$ and
$E(x) \to 1/2$ in the limit of large $N$. For $\ep<3/14$, on the other
hand, a ferromagnetic phase appears, with a fast consensus time
scaling logarithmically with system size, and an exit probability
tending to a step-function.  Finally, in the region $3/14 < \ep <1/4$
there is a mixed phase such that for $|2x-1| <
\sqrt{\frac{3-14\ep}{2\ep-1}}$, we should expect a paramagnetic
disordered phase, while for $|2x-1| > \sqrt{\frac{3-14\ep}{2\ep-1}}$,
the system should order to consensus given by the majority of initial
spins. The boundary separating both behaviors in the mixed phase is
given by an effective $\bar{\ep}(x)$ as
\begin{equation}
  \label{eq:18}
  \bar{\ep}(x) = \frac{x^2-x+1}{2 \left(x^2-x+2\right)},
\end{equation}
such that for $\eps > \bar{\ep}(x)$, a paramagnetic phase is observed,
while $\eps < \bar{\ep}(x)$ corresponds to the ferromagnetic phase.

\begin{figure}[p]
  \begin{center}
    \includegraphics [width=0.75\textwidth]{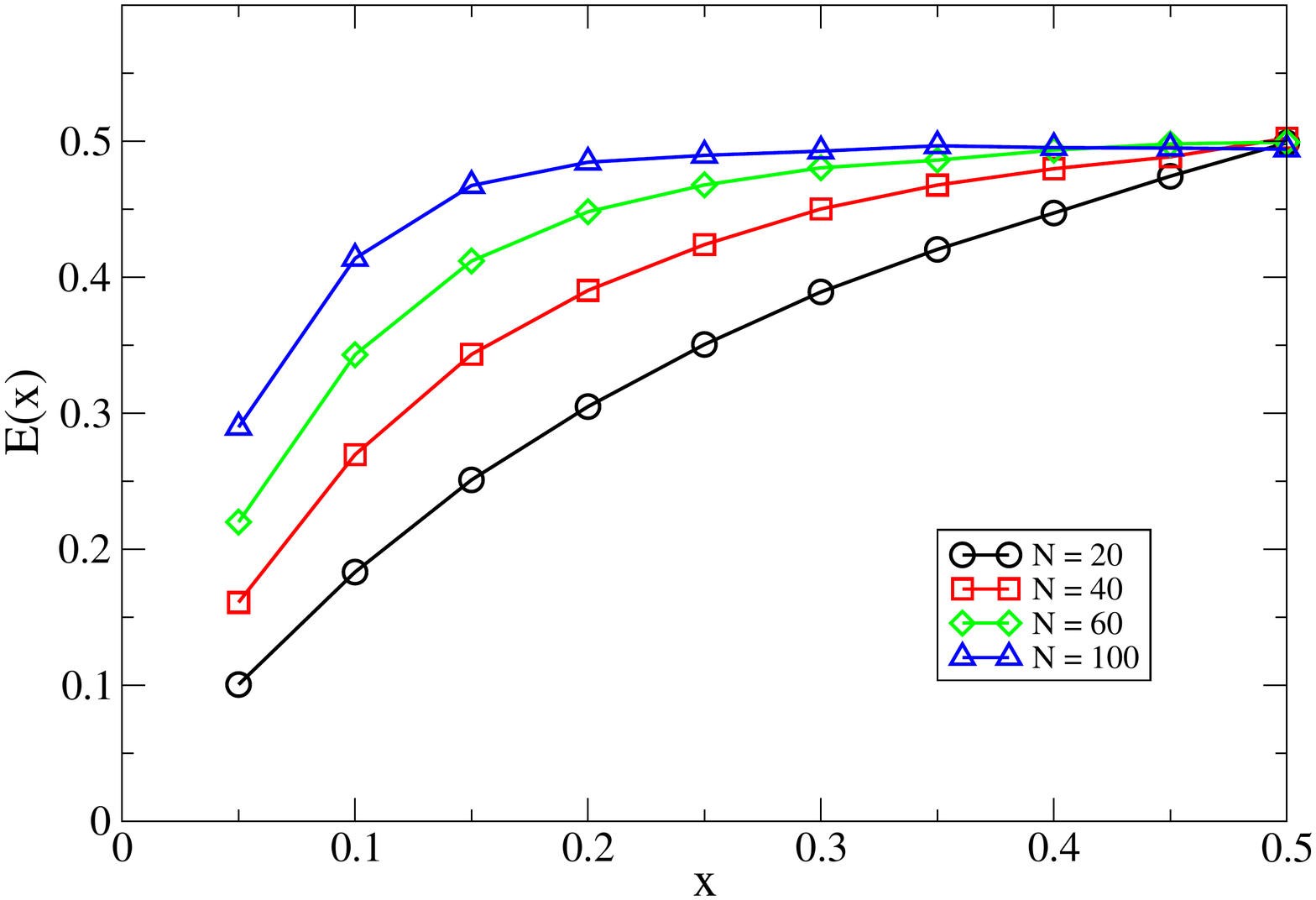}
    \includegraphics [width=0.75\textwidth]{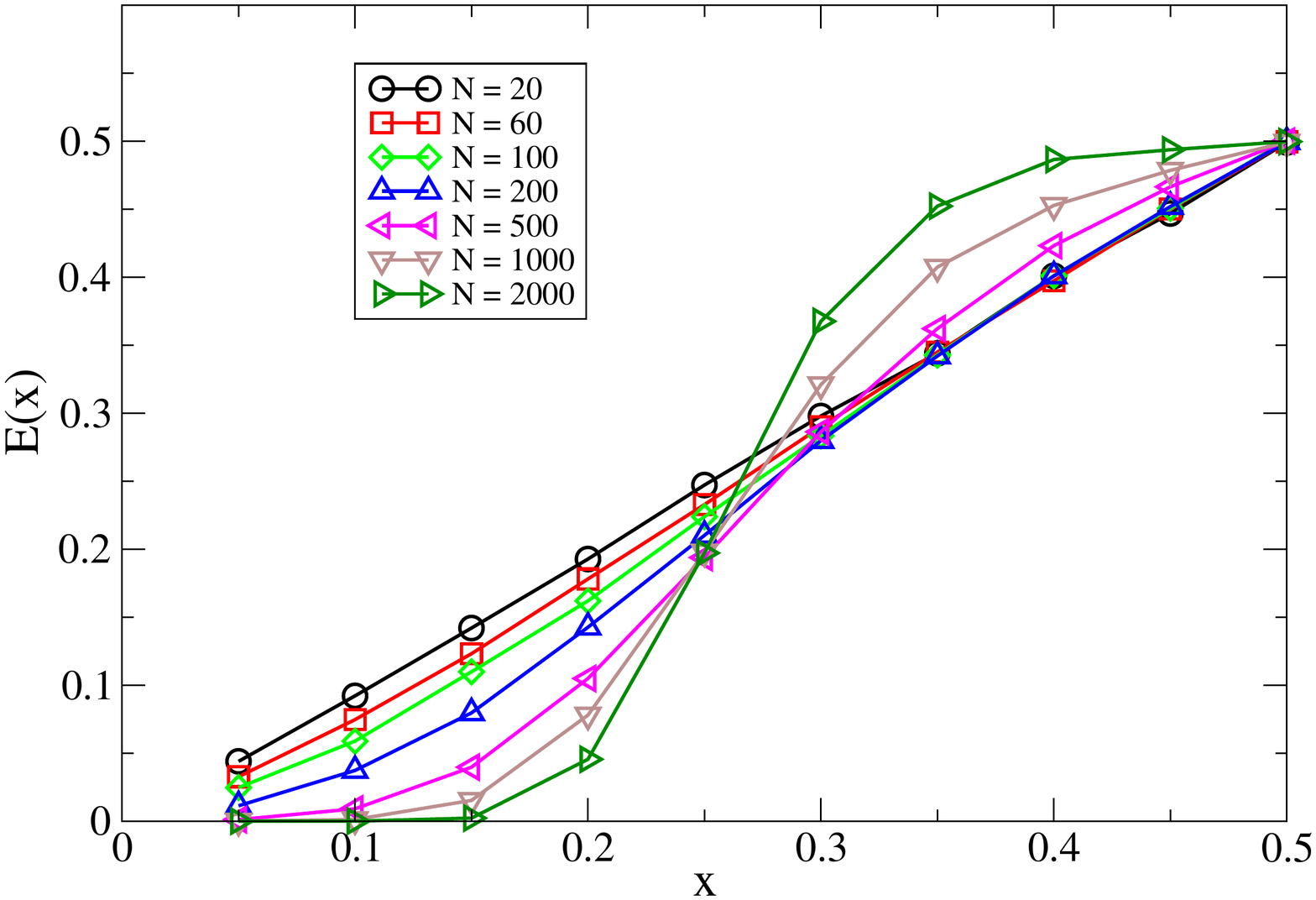}
    \includegraphics [width=0.75\textwidth]{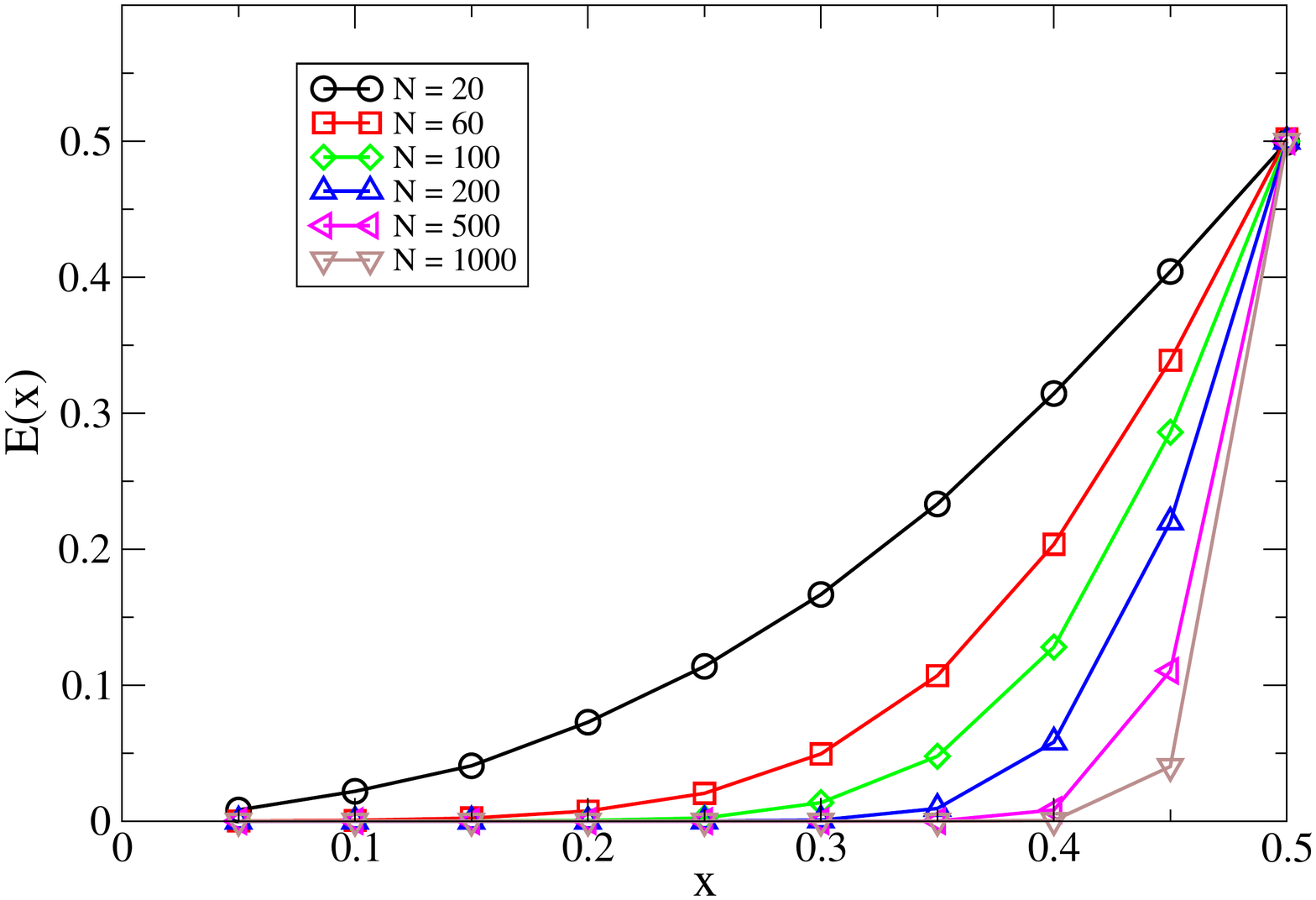}
    \caption{(color online) Numerical check of the behavior of the
      $q=4$ model on fully connected graphs in the different regions
      of the phase-diagram.  The three figures represent the behavior
      of the exit probability for $\eps=0.30$ (top), $\eps=13/58$
      (middle) and $\eps=0.15$ (bottom).  Different curves are for
      increasing values of the system size $N$. {Results averaged
        over $10^5$ independent realizations.}}
\label{fig:q=4behavior}
  \end{center}
\end{figure}
In Fig.~\ref{fig:q=4behavior} we check the predicted behavior in fully
connected networks by measuring numerically the exit probability
$E(x)$ for increasing system size $N$ and three values of $\eps$
representing the three phases discussed above.
For $\eps=0.30>1/4$, $E(x)$ tends to $1/2$, independent
of $x$, as expected in the paramagnetic phase. For $\eps=0.15<3/14$ instead
the tendency is toward a step-function $\Theta(x-1/2)$, indicating
a ferromagnetic phase. In the intermediate regime 
($\eps=13/58=0.2241\ldots$) the ferromagnetic phase dominates for
$x<1/4$ and $x>3/4$, while the region around $x=1/2$ is paramagnetic.

\begin{figure}[t]
  \begin{center}
    \includegraphics [width=0.7\textwidth]{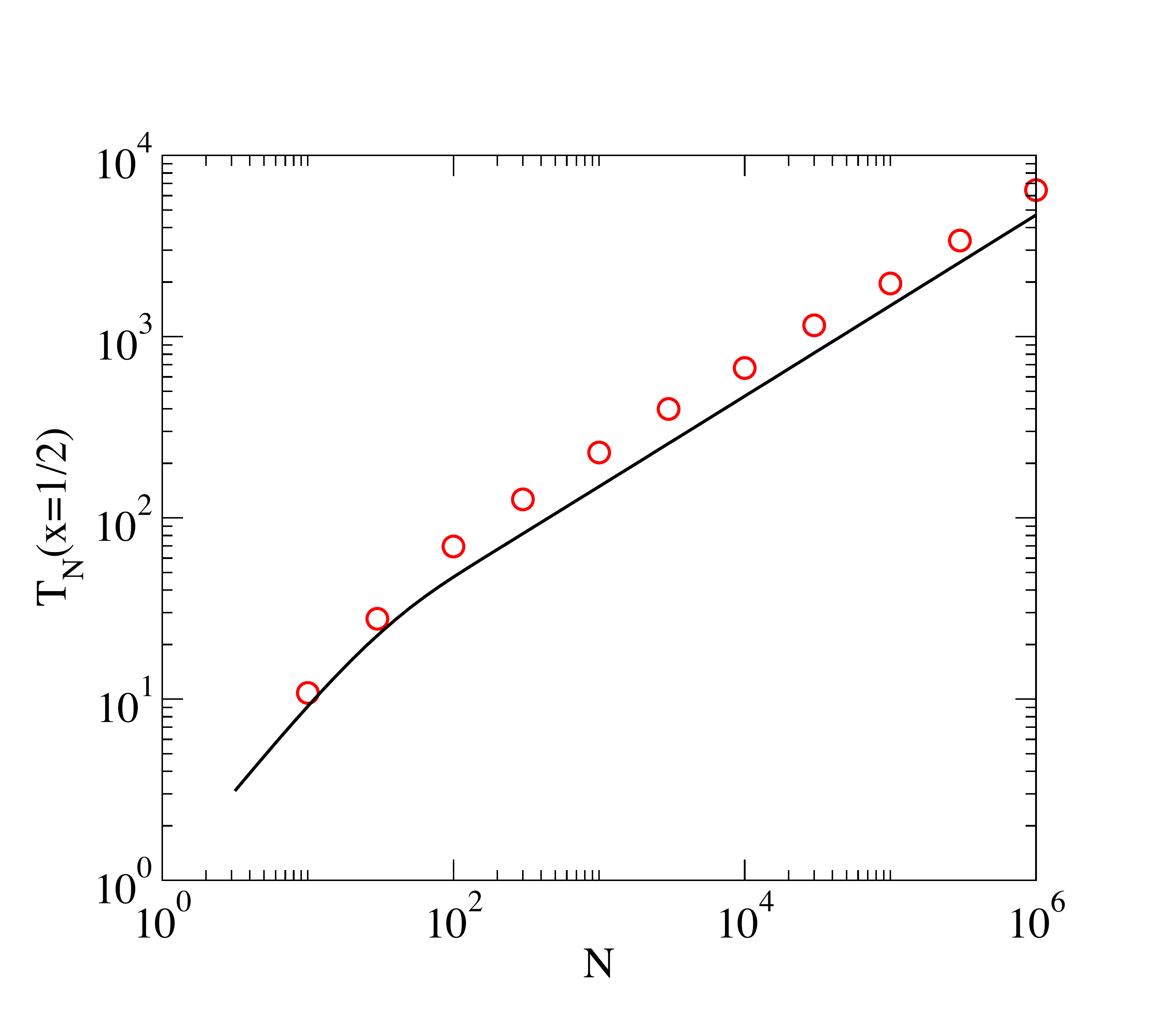}
    \caption{(color online) Size dependence of the consensus time for
      the $q$-voter model on the complete graph for $x=1/2$, $q=4$ and
      $\varepsilon=3/14$.  Numerical simulations of the model
      (circles) are compared with the asymptotic estimate
      Eq.~(\ref{eq:asymptot}) (full line). No fitting parameters are
      introduced. {Numerical results are averaged over at least
        $10^4$ independent realizations.}}
\label{fig:squareroot}
  \end{center}
\end{figure}

In analogy with the GV class, one should expect to observe nontrivial
behavior in the separatrix between the paramagnetic and the
ferromagnetic phase. In the case of the $q=4$-voter model, this
separatrix, $\bar{\ep}(x)$ is not a constant, and therefore it is
impossible to study its behavior in the general case. We shall
therefore focus on the minimum of the separatrix, given by the points
$\eps=3/14$, and $x=1/2$. Considering Eq.~\eqref{eq:6}, and changing
again to the total magnetization variable  $\phi = 2 x-1$, we are led
for $\eps=3/14$ to the equation for the consensus time
\begin{equation}
  \label{eq:20}
  v(\phi) T'_N(\phi) + \frac{1}{2} D(\phi) T''_N(\phi)=-1,
\end{equation}
with boundary conditions reading now $T(\phi=-1) = T(\phi=1) = 0$, and where
\begin{eqnarray}
  \label{eq:22}
  v(\phi) &=&  \frac{1}{14} (1-\phi^2) \phi^3\\
  \label{eq:27}
  D(\phi) &=& \frac{1}{7N} (1-\phi^2)
  (7+6\phi^2)
\end{eqnarray}
By defining
\begin{equation}
  g(\phi)=-\frac{2}{D(\phi)} \;\;\;\mbox{and} \;\;\;
  \mu(\phi)=2\int_0^\phi dz\frac{v(z)}{D(z)}, 
\end{equation}
the exact solution to Eq. (\ref{eq:20}) is given in integral form by
\begin{equation}\label{eq:T_N_MF}
  T_N(\phi)=\int_{-1}^{\phi}dz_1
  e^{-\mu(z_1)}\int_{0}^{z_1} dz_2 \, g(z_2)e^{\mu(z_2)}.
\end{equation}
We notice that since the first exponential is peaked around $z_1=0$ as
$\exp\left(-\frac{Nz_1^4}{28}\right)$, it will select values of
$z_1\approx 0$ in the second integral, which can be thus
expanded for small $z_1$. 
An estimate of the consensus time can thus be expressed in terms of error functions  \cite{abramovitz} as follows
\begin{eqnarray}
  T_N(\phi)&\approx& 
  \sqrt{7\pi N}
  \left[ \mbox{erf}\left(
   \sqrt{\frac{N}{28}} \right) - \mbox{erf}\left(\sqrt{\frac{N\phi^4}{28}} \right) \right]   \nonumber\\
   &-& \frac{2}{3}\left[\exp\left(-\frac{N}{28}\right)-\exp\left(-\frac{N\phi^4}{28}\right)\right].
  \label{eq:asymptot}
\end{eqnarray}
At the point $x=1/2$ ($\phi=0$), the consensus time takes the form 
\begin{equation}
  \label{eq:30}
  T_N(x=1/2) \simeq \sqrt{7\pi N}\;
 \mbox{erf}\left(
   \sqrt{\frac{N}{28}} \right)
 - \frac{2}{3}\left[\exp\left(-\frac{N}{28}\right)-1\right] \sim
 \sqrt{7\pi N} 
\end{equation}
This expression recovers the observed $N^{1/2}$ scaling in the large
$N$ limit \cite{PhysRevE.80.041129}, and is found in good agreement
with numerical simulation results, as shown in
Fig. \ref{fig:squareroot}.

\begin{figure}[t]
  \begin{center}
    \mbox{\includegraphics [width=0.55\textwidth]{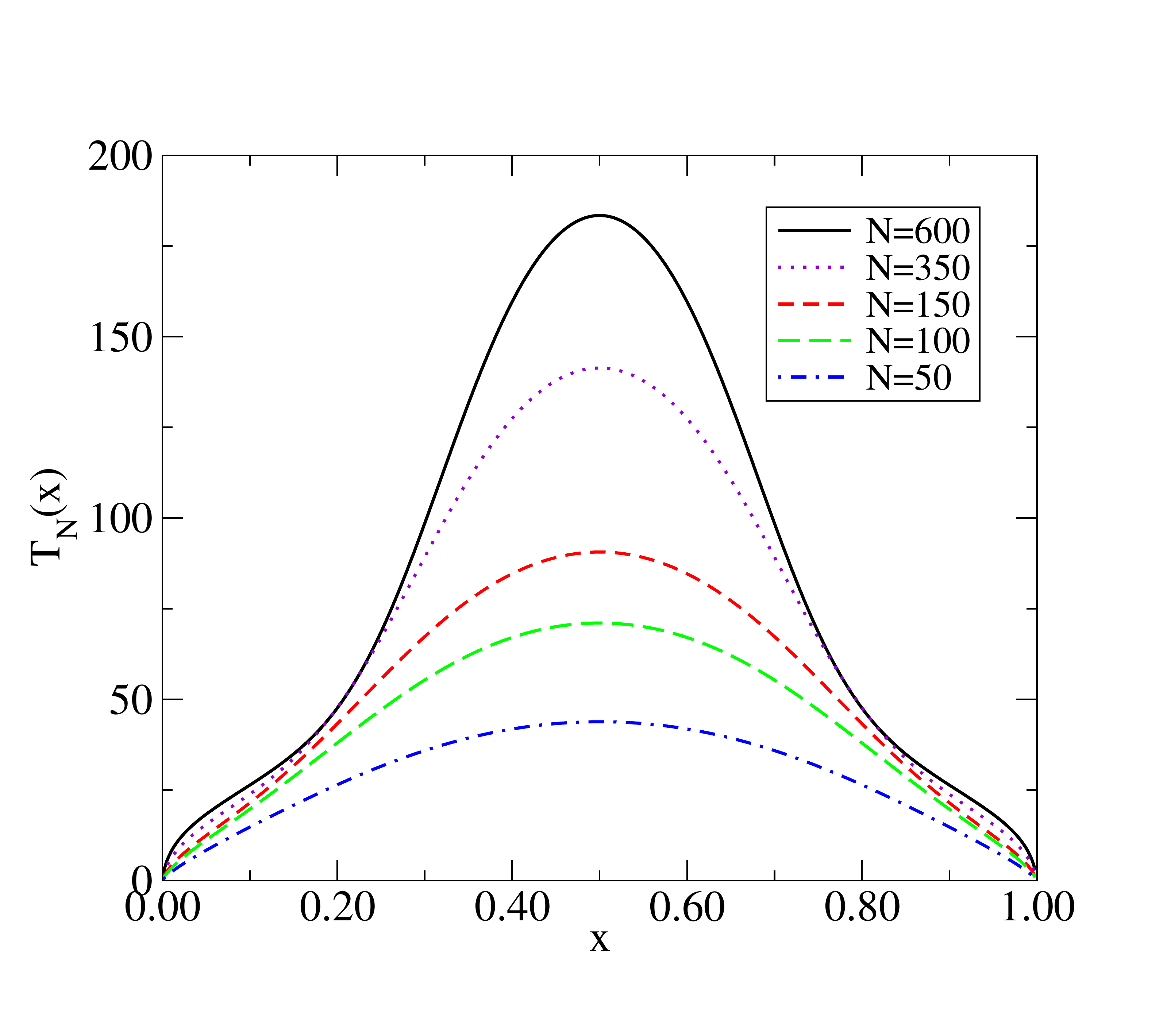}\hspace*{-0.5cm}%
    \includegraphics [width=0.55\textwidth]{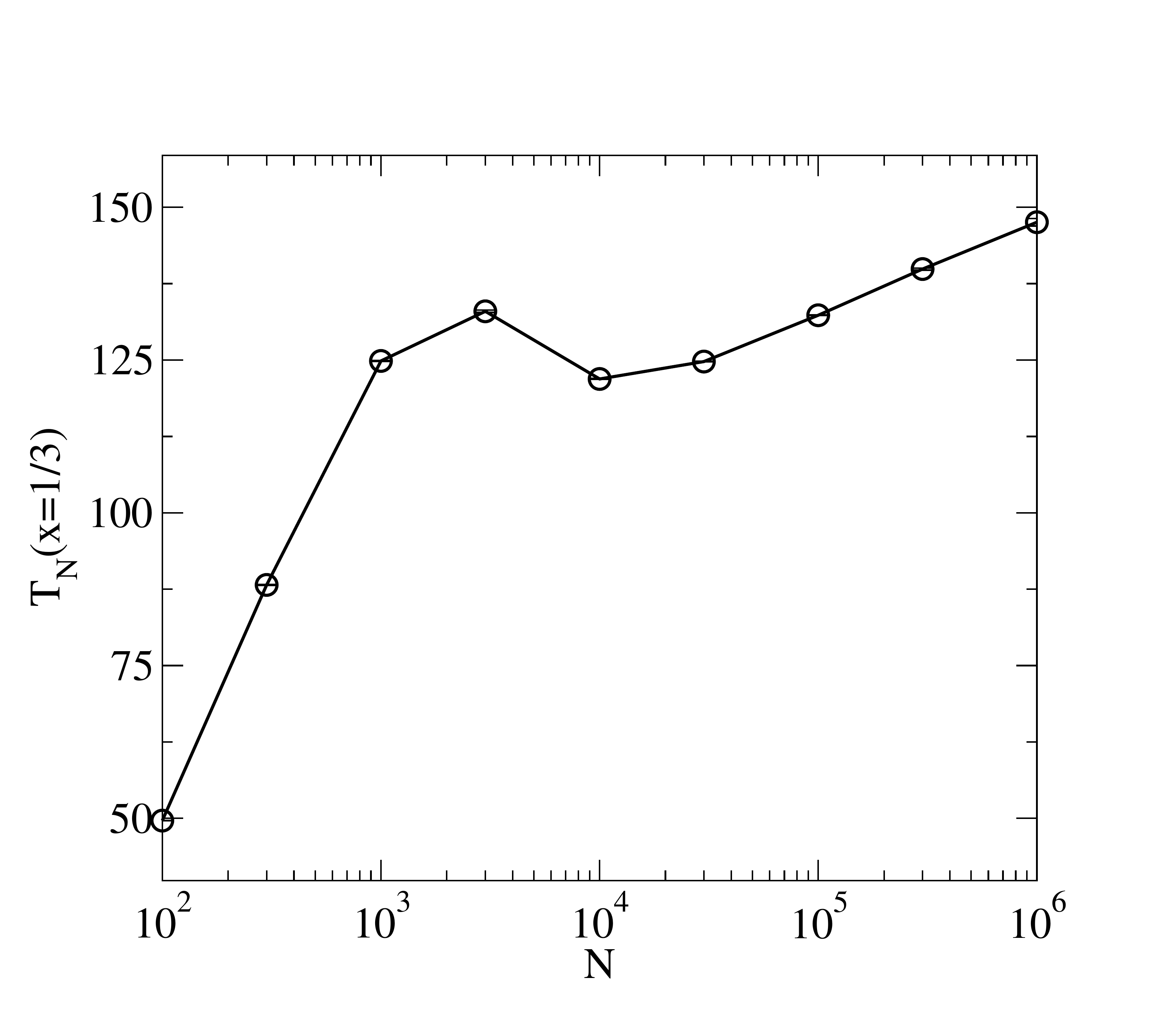}}
  \caption{(color online) Left: Consensus time for the $q$-voter model
    ($q=4$, $\eps=3/14$) on the complete graph as a function of the
    initial condition $x$ for different system sizes. The results of a
    numerical integration of the exact formula Eq.~\eqref{eq:T_N_MF}
    are plotted for increasing values of the system size.
    Right. { Consensus time obtained from numerical simulations as
      a function of $N$ for the initial condition $x=1/3$. Simulation
      results are averaged over at least $10^4$ independent
      realizations.}  The hump in the $T_N$ function is indicative a
    crossover effect.}
    \label{fig:T_N_MF}
  \end{center}
\end{figure}

{For $x \neq 1/2$ the system is in the ferromagnetic phase
and one expects the consensus time to scale logarithmically
for large values of the system size $N$.
By direct numerical integration of Eq.~\eqref{eq:T_N_MF},
the value of the consensus time for any value of $x$ and $N$ can
be computed, see Fig.~\ref{fig:T_N_MF} (left panel). 
The crossover between the behavior at $x=1/2$, $T_N \sim \sqrt{N}$,
and the pure logarithmic growth which asymptotically prevails for
all $x \neq 1/2$, give rise to a nonmonotonicity in the functional
form of $T_N$ as a function of $N$, see Fig.~\ref{fig:T_N_MF}
(right panel). This is a curious phenomenon: there is a short range of
values of $N$ for which the consensus time {\em decreases} as $N$ is
increased.}

\begin{figure}[t]
  \begin{center}
    \includegraphics [width=0.7\textwidth]{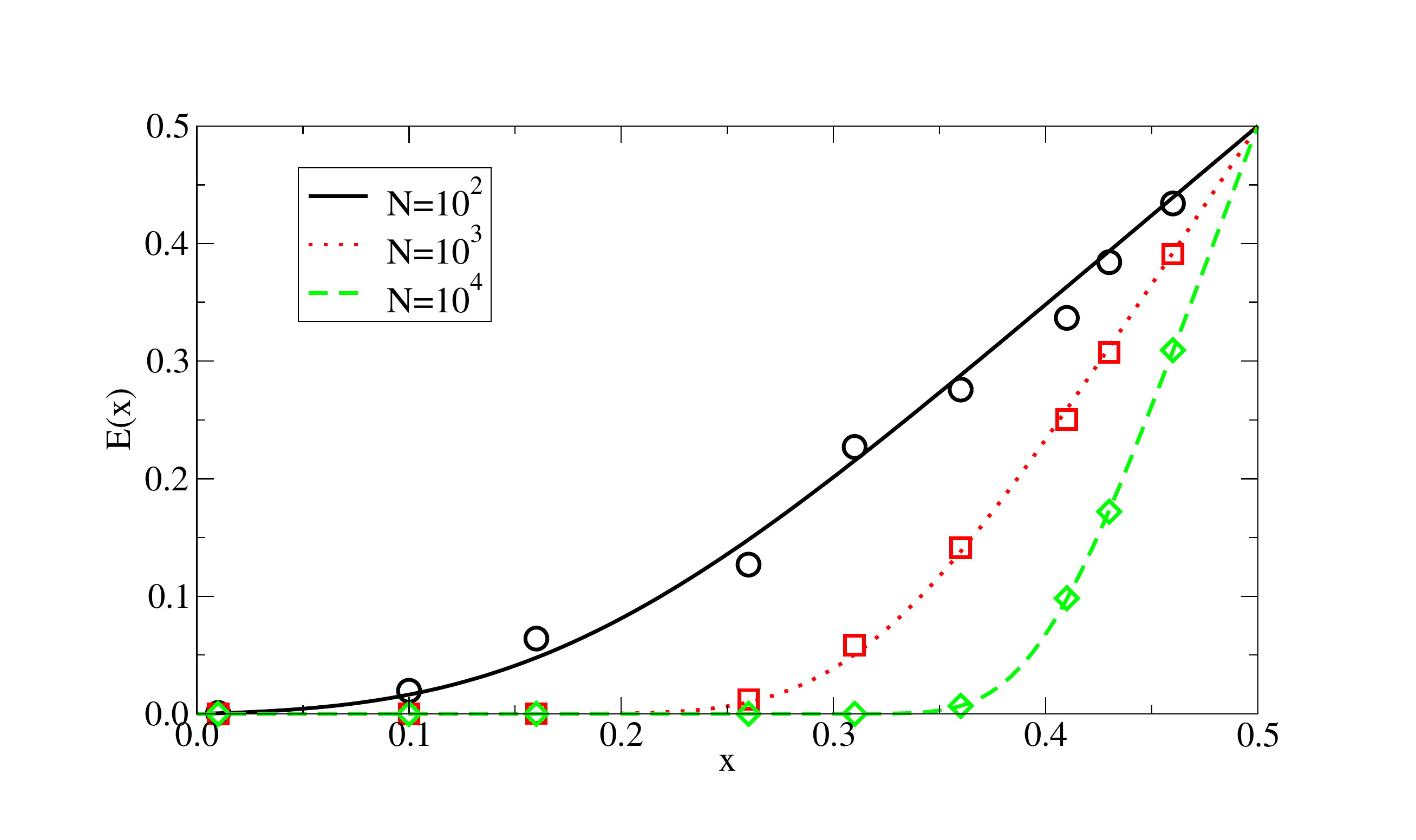}
    \caption{(color online) Exit probability for the $q$-voter model
      (for $q=4$) at the transition point $\eps=3/14$ on fully
      connected networks. {Lines report the theoretical prediction
        Eq.~\eqref{eq:29}. Symbols are the results of numerical
        simulations averaged over $10^4$ realizations of the
        dynamics.} }
\label{fig:exitq4MF}
  \end{center}
\end{figure}

Concerning the exit probability, the solution of the corresponding
equation leads to the form, again expressed in terms of the
magnetization $\phi$,
\begin{equation}
  E(\phi) = \frac{\int_{-1}^\phi e^{-\frac{N}{12} z^2}
    (7+6z^2)^{7N/72} dz}{\int_{-1}^{+1} e^{-\frac{N}{12} z^2}
    (7+6z^2)^{7N/72} dz} 
  \label{eq:4}
\end{equation}
Expanding the integral in powers of $1/N$, and keeping only the first
order, we are led to
\begin{equation}
  \label{eq:28}
  E(\phi) \simeq \frac{\int_{-1}^\phi e^{-z^4 N/28}
    dz}{\int_{-1}^{+1} e^{-z^4 N/28} dz} .
\end{equation}
Focusing on the case $\phi<0$ ($x<1/2$), due to the symmetry of the
expression, the integral can be performed, yielding
\begin{equation}
  \label{eq:29}
  E(\phi) \simeq \frac{1}{2} \frac{\Gamma \left(\frac{1}{4},\frac{N
        \phi^4}{28}\right)-\Gamma
    \left(\frac{1}{4},\frac{N}{28}\right)}{\Gamma 
   \left(\frac{1}{4}\right)-\Gamma \left(\frac{1}{4},\frac{N}{28}\right)}
\end{equation}
where $\Gamma(z)$ is the Gamma function and $\Gamma(\alpha,z)$ is the
incomplete Gamma function \cite{abramovitz}. In
Fig.~\ref{fig:exitq4MF} we plot Eq.~\eqref{eq:29} together with
numerical simulations, displaying again a good agreement for values of
$N$ larger than $100$.

\subsection{Phase diagram for general $q$}
\label{sec:phase-diagr-gener}

\begin{figure}[t]
  \begin{center}
    \includegraphics [width=0.7\textwidth]{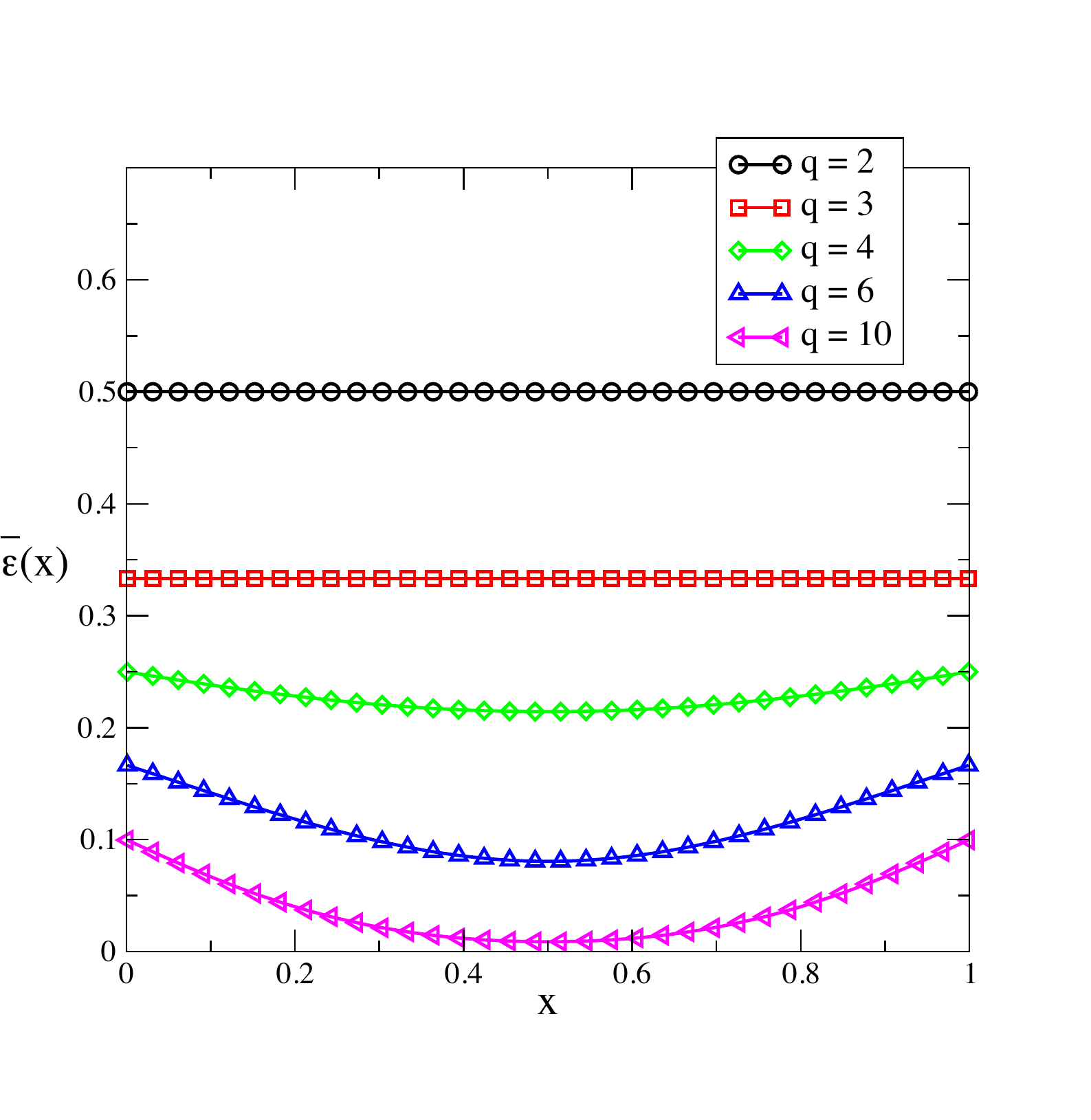}
    \caption{(color online) {Plot of the phase boundary
        $\bar{\eps}(x)$, Eq.~\eqref{eq:33}, as a function of $x$, for
        different values of $q$.}}
\label{fig:boundary}
  \end{center}
\end{figure}

Following the same arguments as in the previous section, we can sketch
a phase diagram for the general case of the $q$-voter model by
considering the sign of the function $\phi v(\phi)$.
Considering the form of the drift in Eq.~\eqref{drift} and the
flipping probability in Eq.~\eqref{f}, simple calculations allow to
show that the quantity  $\phi v(\phi)$ is positive for $\eps <
\bar{\eps}$, with $\bar{\eps}$ taking the form (expressed as a function
of $x$ for simplicity)
\begin{equation}
  \label{eq:33}
  \bar{\eps}(x) = \frac{x(1-x)^q - (1-x)x^q}{(1-2x)[1-x^q -(1-x)^q]}. 
\end{equation}
{In Fig.~\ref{fig:boundary} we plot the shape of the phase
  boundary $\bar{\eps}(x)$, Eq.~\eqref{eq:33}, as a function of $x$,
  for different values of $q$.}
We observe that
\begin{equation}
  \label{eq:34}
  \lim_{x \to \pm 1} \bar{\eps}(x) = \frac{1}{q}, \qquad  
  \lim_{x \to 1/2} \bar{\eps}(x) = \frac{q-1}{2^q-1}.
\end{equation}
We thus obtain that, for the general $q$ case, the $q$-voter model is
in a ferromagnetic phase for $\eps< \frac{q-1}{2^q-1}$, and in a
paramagnetic phase for $\eps>  \frac{1}{q}$. In the region
$\frac{q-1}{2^q-1}<\eps<\frac{1}{q}$, the system is in a mixed
phase whose nature depends on the initial conditions. In the limit of
large $q$, the boundary of region for ferromagnetic behavior decreases
exponentially, while the boundary for paramagnetic behavior decreases
only algebraically.

\section{The $q$-voter model on networks: General heterogeneous
  mean-field theory}
\label{sec:heter-mean-field}

The analytical treatment of voter-like models on complex networks,
beyond the simplest fully connected graph, is traditionally based on the
heterogeneous mean-field (HMF) approach, a powerful tool for the
theoretical analysis of general dynamical processes on heterogeneous
substrates \cite{barratbook,dorogovtsev07:_critic_phenom}. HMF theory
is based on a fundamental approximation: The real (\textit{quenched})
network is coarse-grained into an \textit{annealed} one
\cite{dorogovtsev07:_critic_phenom}, which disregards the specific
connection pattern and postulates that the class of degree $k$ is
connected to the class of degree $k'$ with conditional probability
$P(k'|k)$ \cite{marian1}. 

The application of the HMF formalism to the $q$-voter model follows
from an extension of the case of standard voter model
\cite{PhysRevLett.94.178701,Sood08}, taking into account the intrinsic
multipoint nature of the $q$-voter dynamical rules.  Indeed, the
annealed network approximation is based on the assumption that the
network gets completely rewired between any two dynamical time
steps. In dynamics in which a vertex interacts with a single neighbor,
or with all of them, the numerical (and analytical) implementation
goes by selecting one or $k_i$ (according to the case) neighbors,
randomly chosen with probability $k_j P(k_j)/\av{k}$, where we are
for simplicity making the further assumption
that the network lacks degree correlations,
i.e. $P(k'|k) = k' P(k') / \av{k}$ \cite{Dorogovtsev:2002}. 

In the $q$-voter model, we choose $q$ neighbors, possibly with
repetition, among the $k_i$ neighbors of a vertex. If we choose these
$q$ neighbors at random among all the vertices in the network, we are
implicitly assuming that there is rewiring of the network \textit{even
during the duration of an elementary dynamic step} and this
underestimates the probability that repetitions occur, i.e. a neighbor
is selected multiple times. 
The proper way to eliminate this inconsistency is to select at
random a fixed set of $k_i$ neighbors among all the vertices in the
network, and then to pick up the $q$ participants to the dynamic step
\textit{only} from the fixed set of $k_i$ neighbors previously
selected.

Let us thus consider a generic sparse network.  Assuming no
correlations and annealed approximation, the probability that a
randomly chosen neighbor of a vertex is in state $+1$ is given by
\begin{equation}
  \theta = \sum_{k} \frac{P(k)k}{\langle k\rangle}x_{k},
\end{equation}
where $x_k$ is the probability that a
vertex of degree $k$ is in state $+1$. In the set of $k$
neighbors of a vertex of degree $k$, chosen to make an elementary
update of this vertex, the probability that we choose $n$ in state
$+1$ is given by the binomial distribution $B_{k,n}(\theta)=\binom{k}{n}
\theta^n (1-\theta)^{k-n}$. In this case, if vertex $k$ is in state
$-1$, it will flip with probability
\begin{equation}
  \label{eq:21}
  f_k(n,q) = \left(\frac{n}{k}\right)^q + \eps \left[ 1 -
    \left(\frac{n}{k}\right)^q - \left(\frac{k-n}{k}\right)^q
  \right],
\end{equation}
while if it is in state  $+1$, it will flip with
probability $f_k(k-n,q)$.
Therefore, the probabilities $\Pi(k,\sigma)$ that a microscopic update
will flip a spin of degree $k$ and state $s$
\cite{morettiheterovoter} are given by the elementary
flipping probabilities, $f_k(n,q)$ and $f_k(k-n,q)$, averaged
over the binomial distribution $B_{k,n}(\theta)$, and take the form
\begin{eqnarray}
  \Pi(k,+1)&=&P(k)x_k \sum_{n=0}^k \binom{k}{n} \theta^n
  (1-\theta)^{k-n}
  f_k(n,q),\\
  \Pi(k,-1)&=&P(k)(1-x_k) \sum_{n=0}^k \binom{k}{n} \theta^n
  (1-\theta)^{k-n}
  f_k(k-n,q),
\end{eqnarray}
where the first terms stand for the probability of selecting a vertex
to update of degree $k$, $P(k)$, and the probability that it is in
state $+1$ ($x_k$), or $-1$ ($1-x_k$), respectively. 
The HMF equation for the time evolution of the density $x_k$ takes the form
\begin{equation}
  \label{eq:15}
  \dot{x}_k=  \frac{1}{P(k)}[\Pi(k,-1) - \Pi(k,+1)].
\end{equation}
From this solution, one can find information about the exit
probability by looking for conservation laws of the dynamics in
Eq.~\eqref{eq:15}, while the consensus time follows from an immediate
one-step calculation complemented with an adiabatic approximation,
which allows to replace the multiple degrees of freedom $x_k$ by
single slowly varying conserved quantity
\cite{PhysRevLett.94.178701,Sood08}.

For $q=2$ and any network, the application of this formalism leads to
the trivial result presented in Sec.~\ref{sec:q-voter-model}.  For
values of $q >2$, the application of this formalism is hampered by the
fact that the (possible) conserved quantities in Eq.~\eqref{eq:15} are
not linear functions of $x_k$, which implies that the transformations
leading to the equation for the consensus time cannot be applied
directly \cite{morettiheterovoter}.  We therefore consider in the
following section the simpler case of random regular networks, where a
full analytical solution can be obtained for $q=3$.

\section{The $q$-voter model on random regular networks}
\label{sec:rand-regul-netw}

Random regular networks (RRN) are networks where all nodes have
exactly the same degree $k$, while links are randomly distributed
among them, avoiding self-connections and multiple connections. In
this case, we have $P(k)= \delta_{k, K}$, where $K$ is the average
(constant) degree of the network. In this particular network, since
all vertices share the same value of the degree, a great
simplification arises, as we can set $x_K \equiv \theta \equiv x$,
the total average density of vertices in state $+1$. Additionally,
since the density is homogeneous, we can sidestep the general
formalism from HMF theory and apply directly the Fokker-Planck
approach presented in Sec.~\ref{sec:mean-field-theory}.

In the case $q=2$, the general considerations from
Sec.~\ref{sec:heter-mean-field} indicate the presence of linear voter
behavior for $\eps=1/2$. 

\subsection{Case $q=3$}

In the more interesting case of $q=3$, we are
led, from Eq.~\eqref{eq:21}, \eqref{drift},
and~\eqref{diffusion}, to a drift and a diffusion coefficient of the form
\begin{eqnarray}
  \label{eq:9xxx}
  v(x) &=& \frac{(1-x) x (1-2 x) (K-1)[(3 \eps-1)
   K-1]}{K^2},\\
  \label{eq:10xxx}
  D(x) &=& \frac{(1-x) x \left[3 \eps (K-1) K+K^2 - 2 (K-2)
   (K-1) x(1-x) +1\right]}{K^2 N}.
\end{eqnarray}
The drift turns out now to vanish for a value $\eps_c$ that depends on
$K$ as
\begin{equation}
  \label{eq:11}
  \eps_c(K) = \frac{K+1}{3 K},
\end{equation}
This value tends to  $1/3$, corresponding to a fully
connected network, in the limit $K\to\infty$, but is always larger
than $1/3$ for any finite $K$. This observation indicates that the
structure of a RRN induces more order than a fully connected topology, 
and thus that a larger disordering parameter $\eps$ is needed to
cross over to the disordered paramagnetic phase.  

For this critical value the diffusion coefficient is
\begin{equation}
  \nonumber
  D(x) = \frac{2 (1-x) x \left[K^2-(K-2) (K-1) x(1-x)\right]}{K^2 N}. 
\end{equation}
Using Eq.~\eqref{eq:6} we are led to the equation for the consensus
time
\begin{equation}
  \label{eq:2}
  (1-x)x \left[1-\alpha x(1-x)\right] \partial_x^2 T_N(x) =
  -N,
\end{equation}
where $\alpha = (K-2)(K-1)/K^2$, whose solution is
\begin{eqnarray}
  \label{eq:3}
  T_N(x) &=& -N \left[ x \ln x + (1-x) \ln (1-x)-  \frac{1}{2} \ln
    \left(1 - \alpha x + x^2 \alpha\right) \right. \nonumber \\
  &+& \left.\frac{(1-2 x)
      \sqrt{\alpha } \tan ^{-1}\left(\frac{(1-2 x) \sqrt{\alpha
          }}{\sqrt{4-\alpha }}\right)}{\sqrt{4-\alpha }}-\frac{\sqrt{\alpha } \tan ^{-1}\left(\frac{\sqrt{\alpha
          }}{\sqrt{4-\alpha }}\right)}{\sqrt{4-\alpha }} \right].
\end{eqnarray}
Again, we obtain a non-entropic form, modulated by the factor
$\alpha$, and that yields the mean-field result Eq.~\eqref{eq:9} in
the limit $\alpha\to1$ ($K\to\infty$). 

To check the above results we have simulated the $q=3$-voter model on
RRN with $K=4$. In this case, HMF theory predicts a transition
at
\begin{equation}
  \label{eq:13}
  \eps_c = \frac{5}{12} = 0.4166666.
\end{equation}

\begin{figure}[t]
  \begin{center}
    \includegraphics [width=0.7\textwidth]{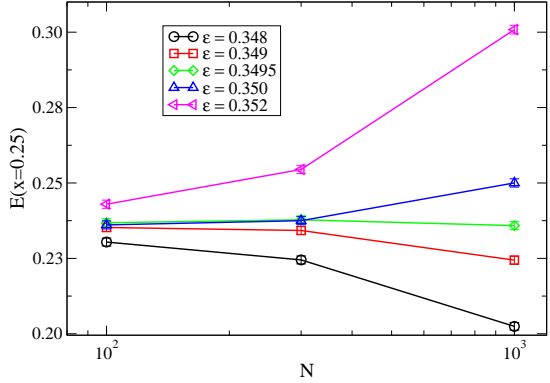}
    \caption{(color online) Exit probability $E(x=0.25)$ for the
      $q$-voter model with $q=3$ on a random regular network with
      $K=4$ as a function of $N$ for different values of
      $\eps$. {The results are obtained by averaging over $10^5$
        numerical simulations of the dynamics.}  The plateau singles
      out the critical value $\eps_c$. }
    \label{E_RRG_q=3_K=4}
  \end{center}
\end{figure}

\begin{figure}[t]
  \begin{center}
    \includegraphics [width=0.7\textwidth]{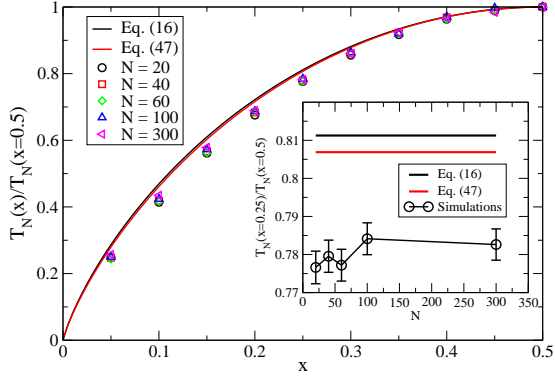}
    \caption{(color online) Main: Normalized consensus time for the
      $q$-voter model with $q=3$ on a random regular network with
      $K=4$, $\eps=\eps_c=0.3495$ and different values of $N$.
      {The numerical results are averaged over $10^5$ realizations
        of the dynamics. Inset: Normalized consensus time for $x=0.25$
        as a function of $N$, showing the disagreement with the
        theoretical predictions.}}
    \label{T_RRG_q=3_K=4}
  \end{center}
\end{figure}

Numerical results lead however at a different, much lower, value.  In
order to estimate numerically the value of $\eps_c$ we have followed
the following approach, based on the behavior of the exit probability
$E(x)$~\cite{Castellano12}.  Indeed, $\eps>\eps_c$ corresponds to a
disordered paramagnetic phase with, for asymptotically large systems,
$E(x) = 1/2$, while $\eps<\eps_c$ corresponds to an ordered
ferromagnetic phase, where $E(x) =\Theta(x)$, the Heaviside theta
function.  Therefore, focusing on an initial density $x<1/2$, we
should observe $E(x) \to 1/2$ for $\eps>\eps_c$, $E(x) \to 0$ for
$\eps<\eps_c$, and $E(x) \to \mathrm{const} < 1/2$ for $\eps=\eps_c$
when increasing the system size $L$. In Fig.~\ref{E_RRG_q=3_K=4} we
report the exit probability for $x=0.25$ and different values of
$\eps$ as a function of the lattice size $N$. A plateau is obtained
for $\eps \simeq 0.3495$, while larger (smaller) values of $\eps$ lead
to an increase (decrease) of $E(x=0.25)$ with $N$. We conclude that
the critical point is in this case located at
$\eps_c=0.3495\pm0.0005$, where the error bars are obtained from the
values of $\eps$ leading to an increasing or decreasing behavior for
$E(x=0.25)$.
Fig.~\ref{E_RRG_q=3_K=4} also indicates that $E(x=0.25)< 0.25$, i.e.
the global form of the exit probability at the transition is not linear,
in agreement with the findings in $d=2$~\cite{Castellano12}.
Not unexpectedly, also the dependence of the consensus time on $x$
(Fig.~\ref{T_RRG_q=3_K=4})
turns out to be different from the analytical prediction, Eq.~(\ref{eq:3})).

\subsection{Case $q=4$}

{The case of $q=4$ in RRN leads to rather complex expressions. In
  particular, as a function of the magnetization $\phi$, the drift
  takes the form
\begin{equation}
  \label{eq:35}
  v(\phi) = \frac{(K-1) \phi  \left(\phi ^2-1\right) \left(14 K^2
   \epsilon -3 K^2+(K-3) (K-2) (2 \epsilon -1) \phi
   ^2+2 K \epsilon -9 K-4 \epsilon +2\right)}{16 K^3}.
\end{equation}
As in the case of the fully-connected graph, no value of $\eps$ can make
this expression zero for any $\phi$. Therefore we exclude again for this
value of $q$ the presence of a linear voter point.
Performing a similar analysis as in Sec.~\ref{sec:case-q=4}, we
can however determine the phase diagram by looking at the sign of the
function $\phi v(\phi)$. It turns out that $\phi v(\phi)$ is positive
for $\eps$ smaller than the threshold $\bar{\eps}$, which, as a
function of $x$, takes the form
\begin{equation}
  \label{eq:36}
  \bar{\eps}(x)=\frac{1}{2}-\frac{(K-2) K}{2 \left(K^2 ((x-1) x+2)+K
   (-5 (x-1) x-1)+6 (x-1) x+1\right)}.
\end{equation}
The phase-diagram is perfectly analogous to the one illustrated
in Fig.~\ref{fig:phasediag}, with a ferromagnetic phase for 
\begin{equation}
 \eps < \lim_{x \to 1/2} \bar{\eps}(x) = \frac{3 K^2+9 K-2}{2 \left(7
      K^2+K-2\right)},
\end{equation}
a paramagnetic phase for
\begin{equation}
\eps >  \lim_{x \to \pm 1} \bar{\eps}(x) = \frac{K^2+K+1}{4 K^2-2 K+2},
\end{equation}
and an intermediate mixed phase, whose nature depends on the initial
condition $x$.

\begin{figure}[t]
  \begin{center}
    \mbox{\includegraphics [width=0.55\textwidth]{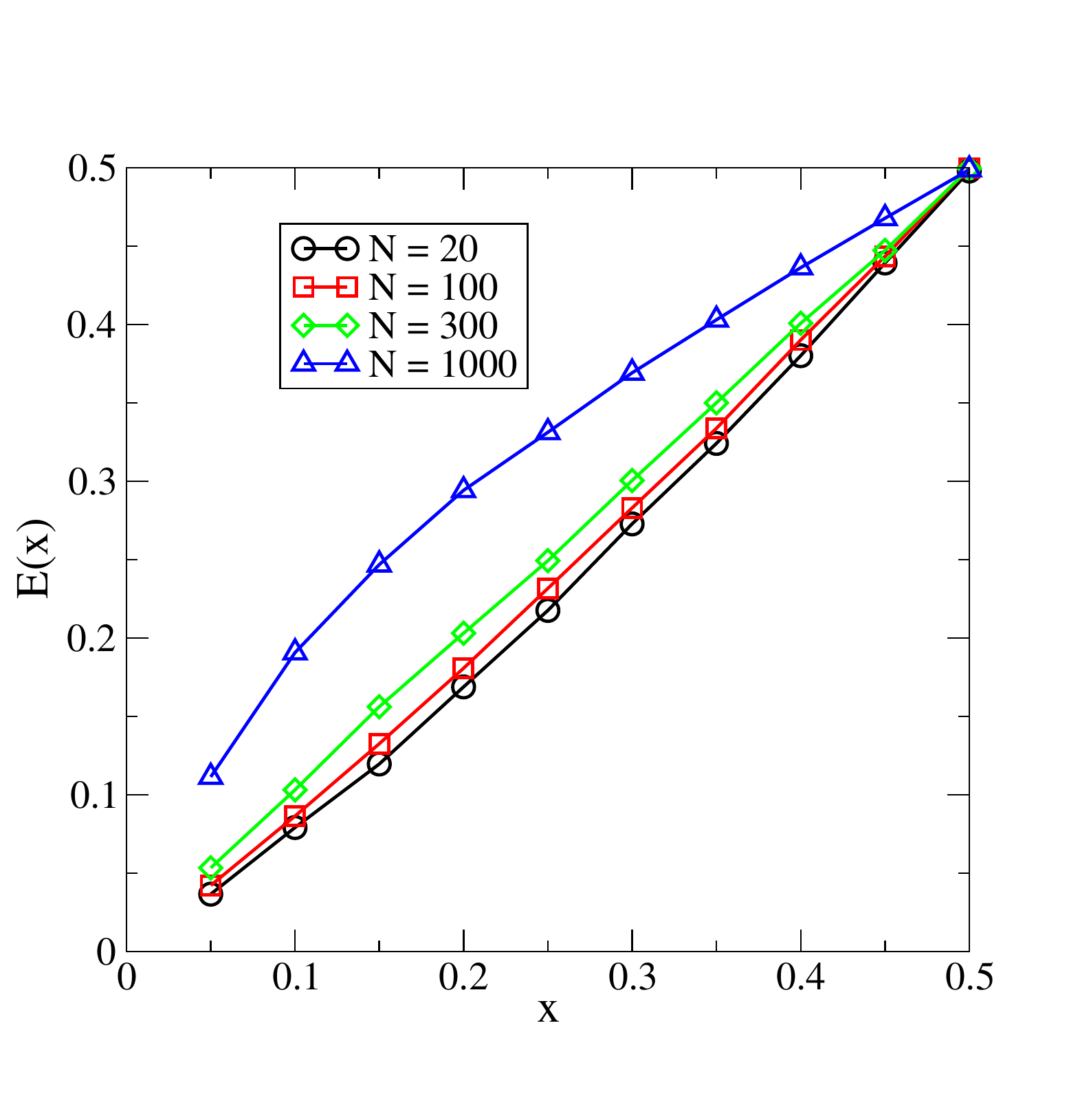}
    \hspace*{-0.5cm}%
    \includegraphics [width=0.55\textwidth]{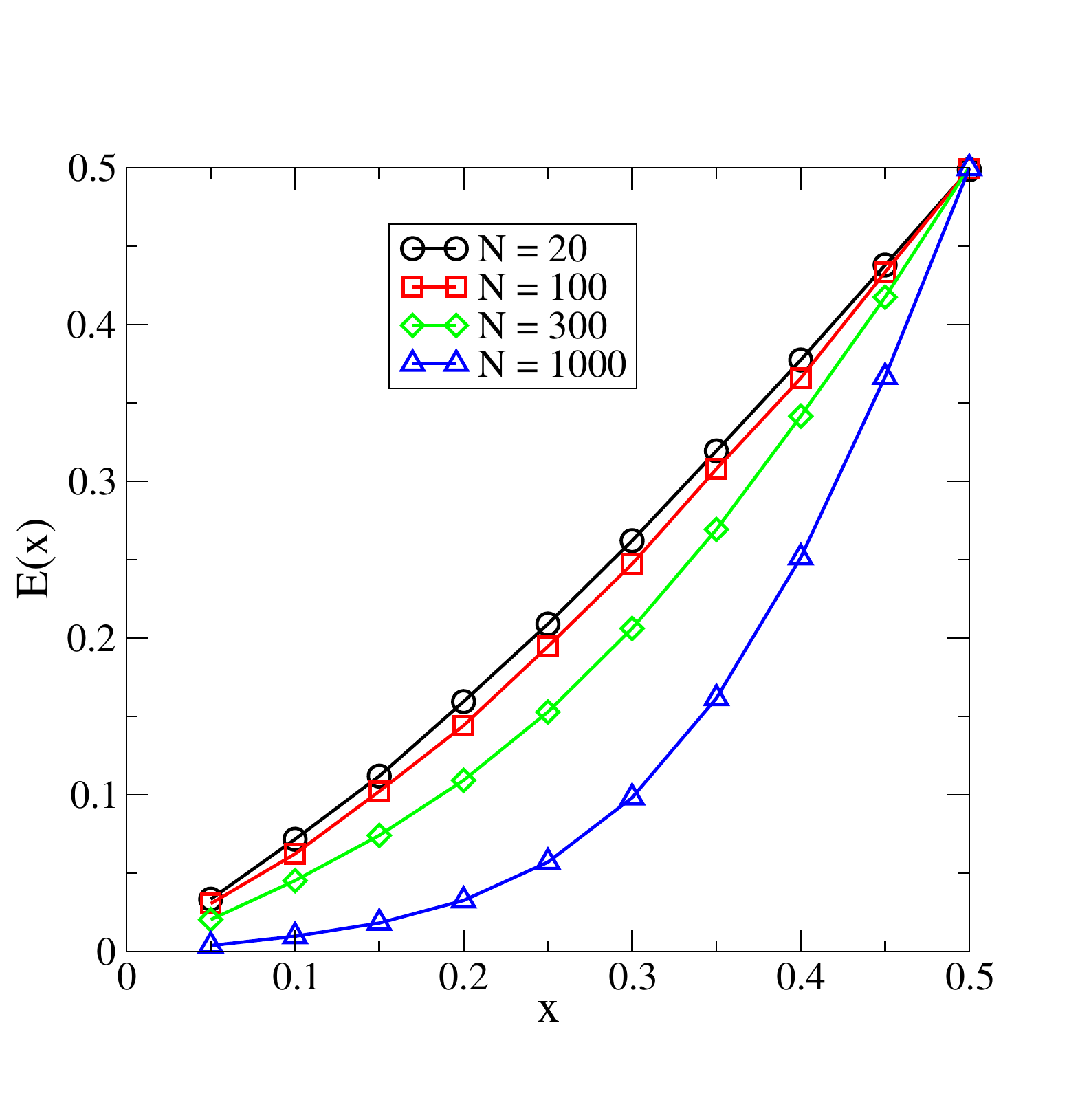}}
  \caption{(color online) Exit probability of the $q$-voter model for
    $q=4$ on random regular networks with $K=4$, slightly above
    ($\eps=0.25$,left) and below ($\eps=0.24$, right) the transition.
    Different symbols are for increasing values of the system size
    $N$.  { It turns out that the system is in the ferromagnetic
      phase for any $x$ for $\eps=0.24$ and in the paramagnetic phase
      for any $x$ for $\eps=0.25$.  The numerical results are averages
      over $10^5$ realizations of the dynamics.}}
    \label{E_RRG_q=4}
  \end{center}
\end{figure}

The interesting question is whether the real $q$-voter dynamics in quenched
RRN networks obeys (at least qualitatively) these theoretical predictions,
in particular with respect to the existence of the mixed phase.
We have checked this by means of numerical simulations in RRN with $K=4$.
In this case, Eq.~(\ref{eq:36}) predicts a mixed phase for values of $\eps$
in the range $21/58 < \eps < 41/114$.
Numerically we observe instead (Fig.~\ref{E_RRG_q=4}) that there is a simple
transition (for $\eps_c \approx 0.247$) separating a purely ferromagnetic
phase for $\eps<\eps_c$ (where $E(x)$ tends to a step-function
$\Theta(x-1/2)$), from a purely paramagnetic phase for $\eps>\eps_c$ 
(where $E(x)$ tends to $1/2$ for any $x$).
Hence we conclude that the fluctuations introduced by the RRN quenched
topology are sufficient to make the behavior of the $q$-voter model
similar to what happens on finite-dimensional
lattices~\cite{PhysRevE.80.041129,Castellano12} and 
qualitatively different from what happens on fully-connected graphs.
We expect similar conclusions to hold for any $q>4$ and for any $K<N-1$.
}

\section{Conclusions}
\label{sec:conclusions}

In this paper we have presented a detailed mean-field analysis
of a model for social dynamics belonging to the generalized
voter universality class. 
A rich phenomenology emerges, in particular for $q \geq 4$,
with a mixed phase separating paramagnetic and ferromagnetic
regions of the phase diagram. When the dynamics occurs on networks
a natural question arising is whether the behavior is of mean-field 
finite dimensional type.
From the analysis of random regular networks it turns out that the
minimal disorder present in those graphs is sufficient to destroy
the mixed phase for $q \geq 4$ and restore a behavior qualitatively
similar to what happens in lattices, with a single transition point
separating ordered and disordered regions of the phase diagram. 
Also in the case $q=3$, the dynamics on RRN is more akin to finite
dimensional lattices (nonlinear exit probability at the transition)
than complete graphs (where $E(x)=x$).
Questions about the behavior on complex, heterogeneous networks
remain open. In that case it is possible that more interesting
phenomena could arise, due to the interplay between the value of $q$
and the degree $k$, which has large fluctuations from node to node.

\section*{Acknowledgments}

R.P.-S.  acknowledges financial support from the Spanish MEC, under
project No. FIS2010-21781-C02-01; the Junta de Andaluc\'{\i}a, under
project No. P09-FQM4682; ICREA Academia, funded by the Generalitat de
Catalunya; partial support by the NSF under Grant No. PHY1066293, and
the hospitality of the Aspen Center for Physics, CO, USA, where part
of this work was performed. P.M acknowledges financial support from
Junta de Andaluc\'{\i}a project P09-FQM4682 and MICINN–FEDER project
FIS2009–08451. S.Y.L acknowledges the support of the 973 Program of
China (No.2012CB720500) and the National High Technology R\&D Program
of China (2012AA041102).


\begin{thebibliography}{10}
\providecommand{\url}[1]{{#1}}
\providecommand{\urlprefix}{URL }
\expandafter\ifx\csname urlstyle\endcsname\relax
  \providecommand{\doi}[1]{DOI~\discretionary{}{}{}#1}\else
  \providecommand{\doi}{DOI~\discretionary{}{}{}\begingroup
  \urlstyle{rm}\Url}\fi

\bibitem{abramovitz}
Abramowitz, M., Stegun, I.A.: Handbook of mathematical functions.
\newblock Dover, New York (1972)

\bibitem{AlHammal05}
{Al Hammal}, O., Chat\'{e}, H., Dornic, I., Mu{\~{ n}}oz, M.A.: Langevin
  description of critical phenomena with two symmetric a bsorbing states.
\newblock Phys. Rev. Lett. \textbf{94}, 230601 (2005)

\bibitem{barabasi02}
Albert, R., Barab\'asi, A.L.: Statistical mechanics of complex networks.
\newblock Rev. Mod. Phys. \textbf{74}, 47--97 (2002)

\bibitem{barratbook}
Barrat, A., Barth\'{e}lemy, M., Vespignani, A.: Dynamical Processes on Complex
  Networks.
\newblock Cambridge University Press, Cambridge (2008)

\bibitem{1751-8121-43-38-385003}
Blythe, R.A.: Ordering in voter models on networks: exact reduction to a
  single-coordinate diffusion.
\newblock J. Phys. A \textbf{43}, 385003 (2010)

\bibitem{blythe07:_stoch_model}
Blythe, R.A., McKane, A.J.: Stochastic models of evolution in genetics, ecology
  and linguistics.
\newblock J. Stat. Mech. p. P07018 (2007)

\bibitem{marian1}
{Bogu\~{n}\'{a}}, M., Pastor-Satorras, R.: Epidemic spreading in correlated
  complex networks.
\newblock Phys. Rev. E \textbf{66}, 047104 (2002)

\bibitem{Bray94}
Bray, A.J.: Theory of phase-ordering kinetics.
\newblock Adv. Phys. \textbf{43}, 357--459 (1994)

\bibitem{Canet05}
Canet, L., Chat\'e, H., Delamotte, B., Dornic, I., Mu\~noz, M.A.:
  Nonperturbative fixed point in a nonequilibrium phase transition.
\newblock Phys. Rev. Lett. \textbf{95}, 100601 (2005)

\bibitem{castellano05:_effec}
Castellano, C.: Effect of network topology on the ordering dynamics of voter
  models.
\newblock AIP Conf. Proc. \textbf{779}, 114 (2005)

\bibitem{Castellano09}
Castellano, C., Fortunato, S., Loreto, V.: Statistical physics of social
  dynamics.
\newblock Rev. Mod. Phys. \textbf{81}, 591--646 (2009)

\bibitem{PhysRevE.80.041129}
Castellano, C., Mu\~noz, M.A., Pastor-Satorras, R.: Nonlinear $q$ -voter model.
\newblock Phys. Rev. E \textbf{80}, 041129 (2009)

\bibitem{Castellano12}
Castellano, C., Pastor-Satorras, R.: Universal and nonuniversal features of the
  generalized voter class for ordering dynamics in two dimensions.
\newblock Phys. Rev. E \textbf{86}, 051123 (2012)

\bibitem{Clifford73}
Clifford, P., Sudbury, A.: {A model for spatial conflict}.
\newblock Biometrika \textbf{60}, 581--588 (1973)

\bibitem{Dornic01}
Dornic, I., Chat\'e, H., Chave, J., Hin~richsen, H.: Critical coarsening
  without surface tension: The universality class of the voter model.
\newblock Phys. Rev. Lett. \textbf{87}, 045701 (2001)

\bibitem{dorogovtsev07:_critic_phenom}
Dorogovtsev, S.N., Goltsev, A.V., Mendes, J.F.F.: Critical phenomena in complex
  networks.
\newblock Rev. Mod. Phys. \textbf{80}, 1275--1335 (2008)

\bibitem{Dorogovtsev:2002}
Dorogovtsev, S.N., Mendes, J.F.F.: Evolution of networks.
\newblock Adv. Phy. \textbf{51}, 1079--1187 (2002)

\bibitem{Drouffe99}
{Drouffe}, J.M., {Godr{\`e}che}, C.: {Phase ordering and persistence in a class
  of stochastic processes interpolating between the Ising and voter models}.
\newblock J. Phys. A \textbf{32}, 249--261 (1999)

\bibitem{Gardinerbook}
Gardiner, C.W.: Handbook of stochastic methods, 2nd edn.
\newblock Springer, Berlin (1985)

\bibitem{Holley:1975fk}
Holley, R.A., Liggett, T.M.: Ergodic theorems for weakly interacting infinite
  systems and voter model.
\newblock Annals of Probability \textbf{3}, 643--663 (1975)

\bibitem{KineticViewRedner}
Krapivsky, P., Redner, S., {Ben-Naim}, E.: A Kinetic View of Statistical
  Physics.
\newblock Cambridge University Press, Cambridge (2010)

\bibitem{liggett99:_stoch_inter}
Liggett, T.M.: Stochastic interacting particle systems: Contact, Voter, and
  Exclusion processes.
\newblock Springer-Verlag, New York (1999)

\bibitem{Molofsky99}
Molofsky, J., Durrett, R., Dushoff, J., Griffeath, D., Levin, S.: Local
  frequency dependence and global coexistence.
\newblock Theoretical Population Biology \textbf{55}, 270 -- 282 (1999)

\bibitem{morettiheterovoter}
{Moretti, P.}, {Liu, S.Y.}, {Baronchelli, A.}, {Pastor-Satorras, R.}:
  Heterogenous mean-field analysis of a generalized voter-like model on
  networks.
\newblock Eur. Phys. J. B \textbf{85}, 88 (2012)

\bibitem{Newman2010}
Newman, M.E.J.: Networks: An introduction.
\newblock Oxford University Press, Oxford (2010)

\bibitem{Deoliveira93}
de~Oliveira, M., Mendes, J., Santos, M.: Nonequilibrium spin models with ising
  universal behaviour.
\newblock J. Phys. A \textbf{26}, 2317--2324 (1993)

\bibitem{Pugliese09}
Pugliese, E., Castellano, C.: Heterogeneous pair approximation for voter models
  on networks.
\newblock Europhys. Lett. \textbf{88}, 58004 (2009)

\bibitem{Sood08}
Sood, V., Antal, T., Redner, S.: Voter models on heterogeneous networks.
\newblock Phys. Rev. E \textbf{77}, 041121 (2008)

\bibitem{PhysRevLett.94.178701}
Sood, V., Redner, S.: Voter model on heterogeneous graphs.
\newblock Phys. Rev. Lett. \textbf{94}, 178701 (2005)

\bibitem{Suchecki05}
Suchecki, K., Egu\'{i}luz, V.M., Miguel, M.S.: Conservation laws for the voter
  model in complex networks.
\newblock Europhys. Lett. \textbf{69}, 228--234 (2005)

\bibitem{Vazquez08}
V\'{a}zquez, F., L\'{o}pez, C.: Systems with two symmetric absorbing states:
  Relating the micro scopic dynamics with the macroscopic behavior.
\newblock Phys. Rev. E \textbf{78}, 061127 (2008)

\end{thebibliography}

\end{document}